\documentclass[11pt,twoside]{article}  % Leave intact
\usepackage{apn3conf}

\begin{document}   % Leave intact

%-----------------------------------------------------------------------
%		            Paper Title 
%-----------------------------------------------------------------------
% Enter the title of the paper.
%
% EXAMPLE: \title{A Breakthrough in Astronomical Software Development}
% 
% If your title is so long as to fill the page header when you print it,
% then please supply a short form as a \titlemark.
%
% EXAMPLE: 
%  \title{Rapid Development for Distributed Computing, with Implications
%         for the Virtual Observatory}
%  \titlemark{Rapid Development for Distributed Computing}
%

\title{X-Ray Observations of Planetary Nebulae}
\titlemark{X-Ray Observations of PNe}

%-----------------------------------------------------------------------
%		          Authors of Paper
%-----------------------------------------------------------------------
% Enter the authors followed by their affiliations.  The \author and
% \affil commands may appear multiple times as necessary (see example
% below).  List each author by giving the first name or initials first
% followed by the last name.  Authors with the same affiliations
% should grouped together. 
%
% EXAMPLE: \author{Margaret Meixner\altaffilmark{1}, Letizia Stanghellini,
%			Howard Bond} 
%          \affil{Space Telescope Science Institute, 
%                 3700 San Martin Dr.,  Baltimore, MD 21218}
%          \author{Joel Kastner}
%          \affil{Rochester Institute of Technology}
%
%          \altaffiltext{1}{Astronomy Department, UIUC}
%
% In this example, the first three authors, "Meixner", "Stanghellini", and
% "Bond" are affiliated with "STScI".  "Meixner" has an alternate 
% affiliation with the "Astronomy Department at UIUC".  The fourth author,
% "Kastner", is affiliated with "Rochester Institute of Technology"

\author{
Mart\'{\i}n A.\ Guerrero\altaffilmark{1}, You-Hua Chu, and Robert A.\ Gruendl}
\affil{
Department of Astronomy, University of Illinois, 1002 W.\ Green St., 
Urbana, IL 61801, USA}
\altaffiltext{1}{
Also at Instituto de Astrof\'{\i}sica de Andaluc\'{\i}a, CSIC, Spain}

%-----------------------------------------------------------------------
%			 Contact Information
%-----------------------------------------------------------------------
% This information will not appear in the paper but will be used by
% the editors in case you need to be contacted concerning your
% submission.  Enter your name as the contact along with your email
% address.
% 
% EXAMPLE:  \contact{Dennis Crabtree}
%           \email{crabtree@cfht.hawaii.edu}
%

\contact{Martin A. Guerrero}
\email{mar@iaa.es}

%-----------------------------------------------------------------------
%		      Author Index Specification
%-----------------------------------------------------------------------
% Specify how each author name should appear in the author index.  The 
% \paindex{ } should be used to indicate the primary author, and the
% \aindex for all other co-authors.  You MUST use the following
% syntax: 
%
% SYNTAX:  \aindex{LASTNAME, F. M.}
% 
% where F is the first initial and M is the second initial (if
% used).  This guarantees that authors that appear in multiple papers
% will appear only once in the author index.  
%
% EXAMPLE: \paindex{Crabtree, D.}
%          \aindex{Manset, N.}        
%          \aindex{Veillet, C.}        
%
% NOTE: this information is also used to build the author list that
% appears in the table of contents.  Authors will be listed in the order
% of the \paindex and \aindex commmands.
%

\paindex{Guerrero, M. A.}
\aindex{Chu, Y.-H.}     % Remove this line if there is only one author
\aindex{Gruendl, R. A.}

%-----------------------------------------------------------------------
%		      Author list for page header	
%-----------------------------------------------------------------------
% Please supply a list of author last names for the page header. in
% one of these formats:
%
% EXAMPLES:
% \authormark{LASTNAME}
% \authormark{LASTNAME1 \& LASTNAME2}
% \authormark{LASTNAME1, LASTNAME2, ... \& LASTNAMEn}
% \authormark{LASTNAME et al.}
%
% Use the "et al." form in the case of seven or more authors, or if
% the preferred form is too long to fit in the header.

\authormark{Guerrero, Chu, \& Gruendl}

%-----------------------------------------------------------------------
%			Subject Index keywords
%-----------------------------------------------------------------------
% Enter up to 6 keywords describing your paper.  These will NOT be
% printed as part of your paper; however, they will be used to
% generate an object index and a subject index for the proceedings.  
% There is no standard list,  however, individual object names are
% encouraged and one or two word descriptions of the topics (e.g.MHD, 
% ionized gas) are useful. 
%
% EXAMPLE:  \keywords{NGC 7027, AFGL 2688, HD 161796, binary stars,
%                      dust,  molecular gas}
%


%-----------------------------------------------------------------------
%			       Abstract
%-----------------------------------------------------------------------
% Type abstract in the space below.  Consult the User Guide and Latex
% Information file for a list of supported macros (e.g. for typesetting 
% special symbols). Do not leave a blank line between \begin{abstract} 
% and the start of your text.

\begin{abstract}          % Leave intact
% Place the text of your abstract here - NO BLANK LINES
Planetary nebulae (PNe) are an exciting addition to the zoo of 
X-ray sources.  
Recent \emph{Chandra} and \emph{XMM-Newton} observations have 
detected diffuse X-ray emission from shocked fast winds in PN 
interiors as well as bow-shocks of fast collimated outflows 
impinging on the nebular envelope.
Point X-ray sources associated with PN central stars are also detected,
with the soft X-ray ($<$0.5 keV) emission originating from the
photospheres of stars hotter than $\sim100,000$~K, and the hard 
X-ray ($\gg$0.5 keV) emission from instability shocks in the
fast stellar wind itself or from a low-mass companion's coronal
activity.
X-ray observations of PNe offer a unique opportunity to directly
examine the dynamic effects of fast stellar winds and collimated
outflows, and help us understand the formation and evolution of PNe.

\end{abstract}

%-----------------------------------------------------------------------
%			      Main Body
%-----------------------------------------------------------------------
% Place the text for the main body of the paper here.  You should use
% the \section command to label the various sections; use of
% \subsection is optional.  Significant words in section titles should
% be capitalized.  Sections and subsections will be numbered
% automatically. 
%
% EXAMPLE:  \section{Introduction}
%           ...
%           \subsection{Our View of the World}
%           ...
%           \section{A New Approach}
%
% It is recommended that you look at the sample papers, sample1.tex
% and sample2.tex, for examples for formatting references, footnotes,
% figures, equations, html links, lists, and other special features.  

\section{Introduction}

Planetary nebulae (PNe) can host different sources of X-ray emission: 
\vspace*{0.025cm}
\begin{enumerate}
  \itemsep0.025cm
\item 
Photospheric emission from hot, 100,000--200,000~K, central stars.  
Such emission is expected at photon energies $\ll$0.5 keV.
\item
Emission from shock-heated gas in PN interiors generated in the 
interaction of the current fast stellar wind (1,000--4,000 
km~s$^{-1}$) with the previous slow AGB wind.  
The shocked fast wind, at temperatures of 10$^7$--10$^8$ K, is too 
tenuous to produce appreciable X-ray emission.
The mixing of nebular shell material into the hot PN interior 
raises the density to produce detectable X-ray emission with a 
limb-brightened morphology.
\item
Emission from shock-heated gas in bow-shocks formed by collimated 
outflows or jets impinging on the AGB wind at velocities $\ge$300 
km~s$^{-1}$.  
The prolonged action of collimated outflows may bore through the
AGB wind and form extended cavities, which can be filled by hot 
shocked gas and emit X-rays, too.
\item
Coronal emission from an unseen and unresolved late-type dwarf 
companion.  
In this case, the PN central star is not responsible for the
X-ray emission.
As stellar coronae have temperatures of a few $\times10^6$~K,
their X-ray emission peaks above 0.5 keV, in sharp contrast to 
the photospheric emission from a hot PN central star.
\end{enumerate}

X-ray observations of hot, shocked-heated gas in PNe allow us to 
examine how fast stellar winds and collimated outflows interact 
with the AGB wind and transfer energy and momentum to the PN 
envelope.
X-ray observations may also reveal unseen faint binary companions 
through their coronal emission, and allow us to assess the importance
of binary shaping of PNe.
Exciting new views of PNe can be obtained through the X-ray window.
In recent years, the \emph{Chandra} and \emph{XMM-Newton X-ray
Observatories} have made major strides in detecting and resolving 
the X-ray emission from PNe.
In this paper we review the X-ray observations of PNe made by these
great X-ray observatories.

\section{X-Ray Observations of PNe}

X-ray emission from PNe was detected in the mid-1980s by \emph{Einstein} 
and \emph{EXOSAT}, but all these detections can be interpreted as soft 
X-ray emission from their hot central stars (see Guerrero, Chu, \& 
Gruendl 2000 for a complete review).  
In the 1990s, \emph{ROSAT} made useful observations of more than 60 PNe; 
only three PNe (A\,30, BD+30$^\circ$3639, and NGC 6543) show marginally 
extended X-ray emission, while two other PNe (NGC\,7009 and NGC\,7293) 
show a hard, $>$0.5 keV, X-ray component not expected from the 
stellar photosphere (Guerrero et al.\ 2000).  
These \emph{ROSAT} observations showed for the first time hints of 
emission from the hot gas in PN interiors, but the evidence 
was not very convincing because of the limited angular resolution 
and low S/N ratios.

\begin{deluxetable}{llll}
%\tablewidth{31.5pc}
\tablecaption{\emph{Chandra} and \emph{XMM-Newton} Detections of PNe}
\tablehead{
\colhead{PN~~~~~~~~~~~~} & 
\colhead{Observatory~~} & 
\colhead{Hard X-ray Emission~~~~~} & 
\colhead{Reference}
}
\startdata 
BD+30$^\circ$3639 & \emph{Chandra}    & Diffuse                  & 1    \\
Hen\,3-1475       & \emph{Chandra}    & Diffuse                  & 2, 3 \\
Mz\,3             & \emph{Chandra}    & Diffuse and Central Star & 4    \\
NGC\,6543         & \emph{Chandra}    & Diffuse and Central Star & 5, 6 \\
NGC\,7009         & \emph{XMM-Newton} & Diffuse                  & 7    \\
NGC\,7027         & \emph{Chandra}    & Diffuse                  & 8    \\
NGC\,7293         & \emph{Chandra}    & Central Star             & 6    
\enddata
\vspace*{-0.55cm}
\tablerefs{
(1) Kastner et al.\ 2000; 
(2) this paper; 
(3) Sahai 2004; 
(4) Kastner et al.\ 2003; 
(5) Chu et al.\ 2001; 
(6) Guerrero et al.\ 2001; 
(7) Guerrero et al.\ 2002; 
(8) Kastner et al.\ 2001.
}
\end{deluxetable}

The launch of modern X-ray observatories, \emph{Chandra} and 
\emph{XMM-Newton}, has made it possible to observe PNe with 
unprecedented sensitivity and angular resolution.  
To date, \emph{Chandra} and \emph{XMM-Newton} have observed 14 PNe.
Diffuse X-ray emission is unambiguously resolved in 6 PNe and hard 
X-ray point sources coincident with the central stars are found in 
3 PNe.  
These detections and their references are listed in Table~1.
The diffuse X-ray sources and hard X-ray point sources are
discussed separately in the next two subsections.

\subsection{Diffuse X-ray Emission from PNe}

The distribution of diffuse X-ray emission relative to the nebular
shell is illustrated for four PNe in Figure~1, where X-ray contours
are overplotted on H$\alpha$ images.
With the exception of Hen\,3-1475, the diffuse X-ray emission from 
each PN is confined within the innermost nebular shell, consistent
with the expectation for shocked fast wind in an 
interacting-stellar-winds model.
% This fact also stands for NGC\,7027 and Mz\,3, not shown here.  
In NGC\,6543 and Mz\,3, the most well-resolved PNe, a limb-brightened 
X-ray morphology can be seen (Chu et al.\ 2001; Kastner et al.\ 2003).

\begin{figure}
\epsscale{1.00}
\plotone{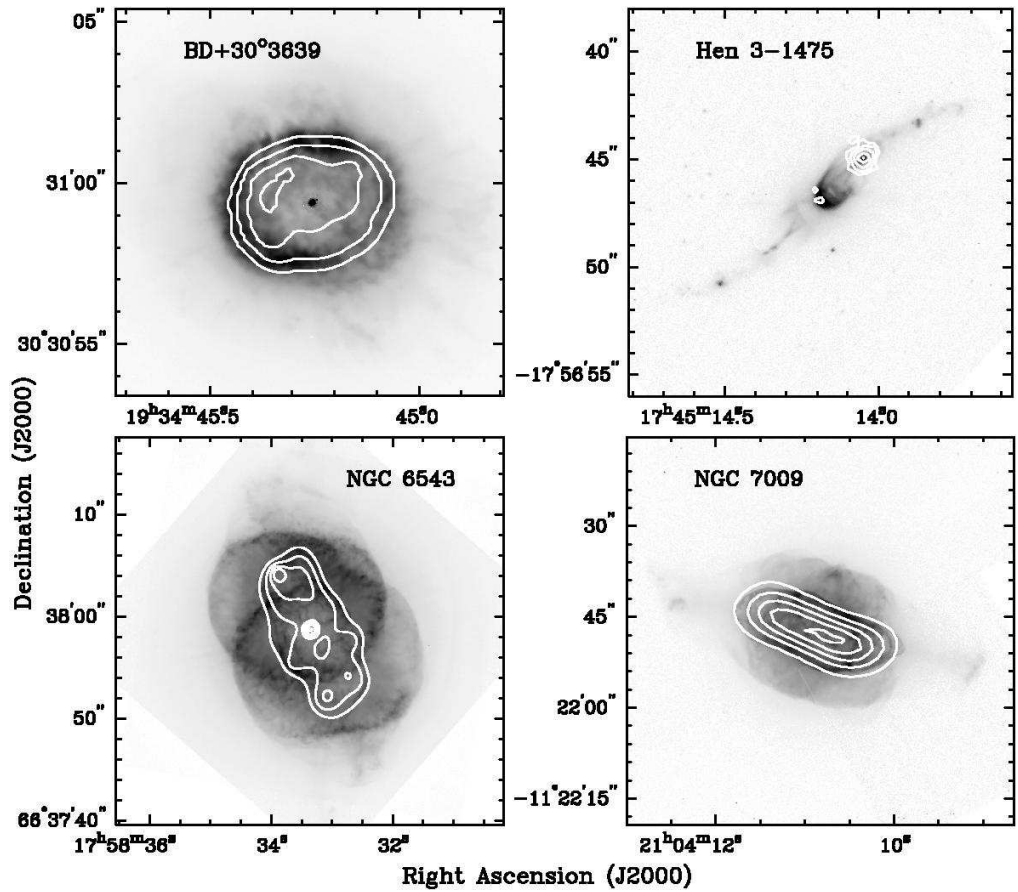}
%\vspace*{9.0cm}
\caption{
\emph{HST} WFPC2 H$\alpha$ images of four PNe with diffuse X-ray 
emission.
The overlaying contours are extracted from X-ray images in the 
0.3-1.0 keV band.  
The X-ray observations of BD+30$^\circ$3639, Hen\,3-1475, and 
NGC\,6543 were obtained with \emph{Chandra} ACIS-S, while NGC\,7009 
was observed with \emph{XMM-Newton} EPIC pn.  
% The angular resolution is $\sim$1$^{\prime\prime}$ and 
% $\sim$6$^{\prime\prime}$ for \emph{Chandra} ACIS-S and \emph{XMM-Newton} 
% EPIC pn, respectively.  
}
\end{figure}

As for Hen\,3-1475, the X-ray emission is located at the tip of a 
bow-shock structure where an abrupt change in velocity of the fast 
collimated outflow emanating from its core has been identified 
(Riera 2004).  
The X-ray emission in Hen\,3-1475 is, thus, associated with its 
collimated outflows, as observed in Herbig-Haro objects (e.g., Pravdo 
et al.\ 2001).  
X-ray-emitting gas shock-heated by collimated outflows has also been 
suggest to exist in Mz\,3 (Kastner et al.\ 2003).  

As most of the X-ray emission from PNe is detected at $<$1 keV (see 
below), this emission is easily absorbed by intervening material,
including the nebular shell and the circumstellar material. 
The importance of internal absorption has been demonstrated
by the anti-correlation between the X-ray surface brightness
and nebular extinction.
Because of the differential absorption across PNe, the X-ray 
morphology of a PN may not be representative of the spatial 
distribution of the hot gas (Kastner et al.\ 2002).  

\begin{figure}
\epsscale{1.00}
\plotone{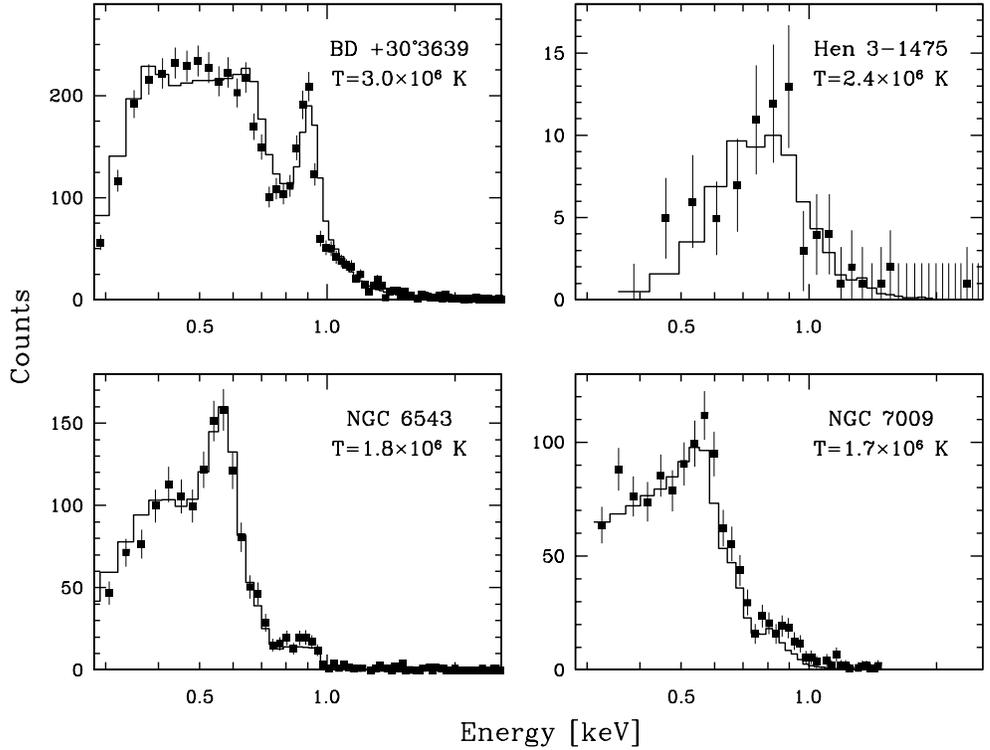}
\caption{
\emph{Chandra} ACIS-S spectra of BD+30$^\circ$3639, Hen\,3-1475, and 
NGC\,6543, and \emph{XMM-Newton} spectrum of NGC\,7009.  
The histogram overlaid on each spectrum corresponds to the best-fit 
model.  
The temperature of this best-fit model is shown in each panel.  
}
\end{figure}

The X-ray spectra of four PNe with diffuse X-ray emission are presented 
in Figure~2.  
Their X-ray emission is soft, peaking at energies $<$1.0 keV.  
The spectral shape is dominated by emission lines of N~{\sc vii}, 
O~{\sc iii}, and Ne~{\sc ix} indicative of thin-plasma emission.  
Spectral fits using a thin-plasma emission model give plasma 
temperatures of 1--3$\times$10$^6$~K and suggest chemical enrichment 
of nitrogen and neon.  
In Hen\,3-1475, the X-ray spectrum implies a hot gas temperature 
corresponds to a shock velocity of $\sim$400 km~s$^{-1}$.

The X-ray luminosities of PNe derived from these spectral fits range 
from 3$\times$10$^{31}$ ergs~s$^{-1}$ to 1$\times$10$^{33}$ ergs~s$^{-1}$.  
The younger PN (BD+30$^\circ$3639, Mz\,3, and NGC\,7027) have 
systematically higher X-ray luminosities and temperatures than the more 
evolved PNe (NGC\,6543 and NGC\,7009).  

\emph{Chandra} observations of another four PNe have resulted in 
non-detections of diffuse X-ray emission.  
In these cases, the PNe either have collimated outflows at
only modest velocities (Hen\,2-90 and M\,1-16), or are evolved
nebulae with no measurable fast winds from the central
stars (NGC\,246 and NGC\,7293).  
The symbiotic star with a bipolar nebula Hen\,2-104 was also not 
detected by \emph{Chandra} observations.  

\subsection{Hard X-ray Emission from PN Central Stars}

The unprecedented \emph{Chandra} resolution has made possible the 
detection of hard X-ray point sources at the central stars of Mz\,3, 
NGC\,6543, NGC\,7293, and possibly Hen\,3-1475 (Guerrero et al.\ 
2001; Kastner et al.\ 2003; this paper).  
Figure~3 shows the X-ray spectra of the central stars of NGC\,6543 
and NGC\,7293.  
These spectra suggest thin plasma emission at temperatures up to 
a few $\times$10$^6$~K and with X-ray luminosities $\sim$10$^{29}$ 
ergs~s$^{-1}$.  

The origin of these point sources is uncertain.  
For the central star of NGC\,7293 (the Helix Nebula), its
temporal  variability in X-rays and in the H$\alpha$ line 
suggests the presence of an unseen dMe companion with an active
corona (Gruendl et al.\ 2001; Guerrero et al.\ 2001).  
In other cases (e.g., NGC\,6543), especially ones with moderate
to strong stellar winds, the instability shocks in the fast stellar
wind itself may be responsible for the hard X-ray emission.

\begin{figure}
\epsscale{1.00}
\plotone{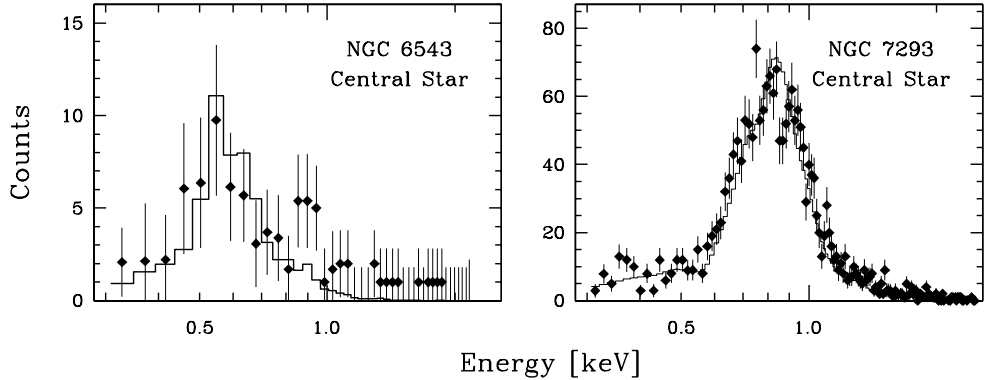}
\caption{
\emph{Chandra} ACIS-S spectra of the central stars of NGC\,6543 and 
NGC\,7293.  
For comparison, the spectrum of a thin plasma emission model with 
temperature of 2$\times$10$^6$~K is overplotted on the X-ray spectrum 
of NGC\,6543.  
The best-fit model with temperature $\sim$7$\times$10$^6$~K is overplotted 
on the X-ray spectrum of NGC\,7293.  
}
\end{figure}

\section{Summary and Future Work}

\emph{Chandra} and \emph{XMM-Newton} observations of PNe have detected
diffuse X-ray emission from hot gas in PN interiors and in bow-shocks 
of fast ($\geq$500 km~s$^{-1}$) collimated outflows, as well as
unresolved point sources at the central stars.
These results have provided a wealth of information on the distribution 
and physical conditions of hot gas in PNe, and allow us to investigate 
the physical structure of PNe as a whole and how collimated outflows 
transfer energy to the nebular envelope.  

The emerging picture revealed by these X-ray observations is that
young PNe with a sharp shell morphology contain significant amounts 
of hot gas in their interiors.  
This hot gas is over-pressurized and drives the nebular expansion,
The duration for the presence of hot gas is short, as only the
youngest PNe have detectable diffuse X-ray emission.
It is possible that excessive mixing of nebular material lowered
the hot gas temperatures to below $1\times10^6$~K, where the 
cooling function peaks and a runaway cooling ensues.

\emph{Chandra} and \emph{XMM-Newton} have the ideal resolution 
and sensitivity to observe PNe.  
As the amounts of hot gas in PN interiors or bow-shocks of 
collimated outflows are usually small, PNe are faint X-ray
sources.  
One must be careful in selecting PN targets for X-ray observations.
Factors that should be taken into consideration include: fast
stellar wind strength, speed of collimated outflows, foreground 
absorption, nebular shell morphology, etc.
As presented by Chu et al.\ (2004), the O~{\sc vi} 
$\lambda\lambda$1032,1037 lines provide a promising diagnostic
for the existence of 10$^6$~K hot gas for PNe with central stars
cooler than 125,000~K.
While we need to actively request new \emph{Chandra} and 
\emph{XMM-Newton} observations of PNe, the targets must be 
carefully selected with the above considerations to maximize
the likelihood of detection.
Only positive detections can be analyzed and advance our
understanding of physical structures of PNe.

%-----------------------------------------------------------------------
%			      References
%-----------------------------------------------------------------------
% List your references below within the reference environment
% (i.e. between the \begin{references} and \end{references} tags).
% Each new reference should begin with a \reference command which sets
% up the proper indentation.  Observe the following order when listing
% bibliographical information for each reference:  author name(s),
% publication year, journal name, volume, and page number for
% articles.  Note that many journal names are available as macros; see
% the User Guide listing "macro-ized" journals.   
%
% EXAMPLE:  \reference Hagiwara, K., \& Zeppenfeld, D.\  1986, 
%                Nucl.Phys., 274, 1
%           \reference H\'enon, M.\  1961, Ann.d'Ap., 24, 369
%           \reference King, I.\ R.\  1966, \aj, 71, 276
%           \reference King, I.\ R.\  1975, in Dynamics of Stellar 
%                Systems, ed.\ A.\ Hayli (Dordrecht: Reidel), 99
% 
% Note the following tricks used in the example above:
%
%   o  \& is used to format an ampersand symbol (&).
%   o  \'e puts an accent agu over the letter e.  See the User Guide
%      and the sample files for details on formatting special
%      characters.  
%   o  "\ " after a period prevents LaTeX from interpreting the period 
%      as an end of a sentence.
%   o  \aj is a macro that expands to "Astron. J."  See the User Guide
%      for a full list of journal macros
%

% Do not place any material after the references section


\begin{references}

\reference Chu, Y.-H., Gruendl, R.\ A., \& Guerrero, M.\ A.\ 2004, 
   in this volume

\reference Chu, Y.-H., Guerrero, M.\ A., Gruendl, R.\ A., Williams, R.\ M., 
\& Kaler, J.\ B.\ 2001, \apj, 553, L69

\reference Gruendl, R.\ A., Chu, Y.-H., O'Dwyer, I., \& Guerrero, M.\ 
A.\ 2001, \aj, 122,~308

\reference Guerrero, M.\ A., Chu, Y.-H., \& Gruendl, R.\ A.\ 2000, 
\apjs, 129, 295

\reference Guerrero, M.\ A., Chu, Y.-H., \& Gruendl, R.\ A.\ 2002, 
\aap, 387, L1

\reference Guerrero, M.\ A., Chu, Y.-H., Gruendl, R.\ A., Williams, R.\ M., 
\& Kaler, J.\ B.\ 2001, \apj, 553, L55

\reference Kastner, J.\ H., Balick, B., Blackman, E.\ G.,et al.\
    2003, \apjl, 591, L37

\reference Kastner, J.\ H., Li, J., Vrtilek, S.\ D., et al.\ 2002,
   \apj, 581, 1225

\reference Kastner, J.\ H., Soker, N., Vrtilek, S.\ D., \& Dgani, R.\ 
2000, \apj, 545, L57

\reference Kastner, J.\ H., Vrtilek, S.\ D., \& Soker, N.\ 2001, \apj, 
550, L189

\reference Pravdo, S.\ H., Feigelson, E.\ D., Garmire, G., et al.\
   2001, Nature, 413, 708 

\reference Riera, A.\ 2004, in this volume

\reference Sahai, R.\ 2004, in this volume

\end{references}
\end{document}